%% file: vfdm.tex
\documentclass[a4paper,12pt]{article}
\usepackage[latin1]{inputenc}
\usepackage{amsmath}
\usepackage{epsfig}
\usepackage{amssymb}

\input{macros}

\title{Vandermonde Frequency Division Multiplexing for Cognitive Radio}

\author{
\begin{footnotesize}
\centering
\begin{tabular}{cc}
	Leonardo S. Cardoso and Mari Kobayashi and M\'erouane Debbah & {\O}yvind Ryan\\
	SUPELEC & University of Oslo\\
	Gif-sur-Yvette, France & Oslo, Norway\\
	\{leonardo.cardoso,mari.kobayashi,merouane.debbah\}@supelec.fr & oyvindry@ifi.uio.no\\
\end{tabular}
\end{footnotesize}
}

\begin{document}

\maketitle

\begin{abstract}
We consider a cognitive radio scenario where a primary and a secondary user wish to communicate with their corresponding receivers simultaneously over frequency selective channels. Under realistic assumptions that the secondary transmitter has no side information about the primary's message and each transmitter knows only its local channels, we propose a Vandermonde precoder that cancels the interference from the secondary user by exploiting the redundancy of a cyclic prefix. Our numerical examples show that VFDM, with an appropriate design of the input covariance, enables the secondary user to achieve a considerable rate while generating zero interference to the primary user.
\end{abstract}

\section{Motivation}

\begin{figure}[t]
 \centering
 \includegraphics[width=0.4\columnwidth]{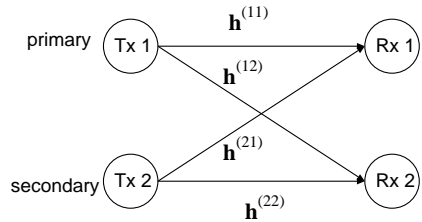}
 \caption{$2\times 2$ cognitive model}
 \label{fig:Model}
\end{figure}

We consider a $2\times 2$ cognitive radio model where both a primary
(licensed) transmitter and a secondary (unlicensed) transmitter wish
to communicate with their corresponding receivers simultaneously as
illustrated in Fig.\ref{fig:Model}. When both transmitters do not
share each other's message, the information theoretic model falls
into the interference channel
\cite{carleial-interference-1978,sato-multiuserchannels-1977} whose
capacity remains open in a general case. A significant number of
recent works have aimed at characterizing the achievable rates of
the cognitive radio channel, i.e. the interference channel with some
knowledge of the primary's message at the secondary transmitter
\cite{devroye2006arc,jovicic-cognitiveradio-2006,maric2007cic,maric-interferencechannel-2007}. 
These include the pioneering work of \cite{devroye2006arc}, the
works of
\cite{jovicic-cognitiveradio-2006},  
\cite{maric2007cic} for the case of weak, strong Gaussian
interference, respectively, and finally a recent contribution of
\cite{maric-interferencechannel-2007} with partial knowledge at the
secondary transmitter. In all these works, the optimal transmission
scheme is based on dirty-paper coding that pre-cancels the known
interference to the secondary receiver and helps the primary user's
transmission. Unfortunately, this optimal strategy is very complex
to implement in practice and moreover based on rather unrealistic
assumptions : a) the secondary transmitter has full or partial
knowledge of the primary message, b) both transmitters know all the
channels perfectly. Despite its cognitive capability, the assumption
a) seems very difficult (if not impossible) to hold. This is because
in practice the secondary transmitter has to decode the message of
the primary transmitter perfectly in a causal manner by training
over a noisy, faded or capacity-limited link. The assumption b)
requires both transmitters to perfectly track all channels (possibly
by an explicit feedback from two receivers) and thus might be
possible only if the underlying fading channel is quasi-static.

The above observation motivates us to design a practical
transmission scheme under more realistic assumptions. First, we
consider no cooperation between two transmitters. The
primary user is ignorant of the secondary user's presence and
furthermore the secondary transmitter has no knowledge on the
primary transmitter's message. Second, we assume that each
transmitter $i$ knows perfectly its local channels ($\hv^{(i1)}$ and
$\hv^{(i2)}$) and each receiver $i$ knows only its direct channel
($\hv^{(ii)}$). This assumption is rather reasonable when the
channel reciprocity can be exploited under time division duplexing systems. Finally,
assuming frequency selective fading channels, we consider OFDM transmission. The last
assumption has direct relevance to the current OFDM-based standards such as WiMax,
802.11a/g, LTE and DVB \cite{standard802.11}. 
Under this setting, there is clearly a tradeoff between the
achievable rates of the two users. For the cognitive radio application, however,
one of the most important goals is to design a transmit scheme of the
secondary user that generates zero interference to the primary receiver.

We propose a linear Vandermonde precoder that generates zero
interference at the primary receiver by exploiting the redundancy of
a cyclic prefix and name this scheme {\it Vandermonde Frequency
Division Multiplexing} (VFDM). The precoder exploits the frequency
selectivity of the channel rather than the spatial dimension and can
be considered as a frequency beamformer (in comparison to the
classical spatial beamformer). More precisely, our precoder is given
by a Vandermonde matrix \cite{golub1996mc} with $L$ roots
corresponding to the channel $\hv^{(21)}$ from the primary user to
the secondary receiver. The orthogonality between the precoder and
the channel enables the secondary user to send $L$ symbols,
corresponding to the size of a cyclic prefix, while maintaining zero
interference. This is contrasted with the approach of
\cite{jovicic-cognitiveradio-2006} where the zero interference is
limited to the case of weak interference. To the best of our
knowledge, a Vandermonde precoder to cancel the interference has
never been proposed. The use of the Vandermonde filter together with
a Lagrange spreading code was proposed to cancel the multiuser
interference on the uplink of a CDMA system \cite{scaglione2000lvm}.
However, this scheme is conceptually different in that its
interference cancellation exploits the orthogonality between the
spreading code and the filter. Moreover, it does not depend on the
channel realization.

Since the size $L$ of the cyclic prefix is typically fixed to be
much smaller than the number $N$ of OFDM symbols (sent by the
primary user) \cite{standard802.11}, VFDM is highly
suboptimal in terms of the achievable rate. 
Nevertheless, we show
that the secondary user can improve its rate by 
appropriately designing its input covariance
at the price of additional side information on the interference plus noise covariance seen by the secondary receiver.
Numerical examples inspired by IEEE 802.11a setting show that 
VFDM with our proposed covariance design enables the secondary user
to achieve a non-negligible rate of 8.44 Mbps while guaranteeing the primary user 
its target rate of 36 Mbps with the operating SNR of 10 dB. 
Finally, although this paper focuses on the zero interference case desired for the
cognitive radio application, VFDM can be suitably
modified to provide a tradeoff between the amount of interference
that the secondary transmitter cancels and the rate that it
achieves. We discuss in Section \ref{sect:InputCovariance} some
practical methods to generalize VFDM.


\section{System Model}

We consider a $2\times 2$ cognitive model in Fig.\ref{fig:Model}
over frequency selective fading channels. By letting $\hv^{(ij)}$
denote the channel with $L+1$ paths between transmitter $i$ and
receiver $j$, we assume that entries of $\hv^{(ij)}$ are i.i.d.
Gaussian $\sim\Nc_{\Cc}(0, \sigma_{ij}/(L+1))$ and moreover the
channels are i.i.d. over any $i,j$. In order to avoid
block-interference, we apply OFDM with $N$ subcarriers with a cyclic
prefix of size $L$. The receive signal for receiver 1 and receiver 2
is given by
\begin{eqnarray} \label{OriginalOutput}\nonumber
    \yv_1 &=& \Fm \left(\Tc(\hv^{(11)})\xv_1 + \Tc(\hv^{(21)})\xv_2+ \nv_1 \right)\\ 
    \yv_2 &=& \Fm \left(\Tc(\hv^{(22)}) \xv_2 + \Tc(\hv^{(12)}) \xv_1+ \nv_2 \right)
\end{eqnarray}
where $\Tc(\hv^{(ij)})$ is a $N \times (N+L)$ Toeplitz with vector $\hv^{(ij)}$
\begin{small}
\begin{eqnarray*}
\Tc(\hv^{(kj)}) =
\left[ \begin{array}{cccccc}
h_{L}^{(kj)} & \cdots & h_0^{(kj)} & 0 & \cdots & 0 \\
0 & \ddots &  & \ddots & \ddots & \vdots \\
\vdots & \ddots & \ddots &  & \ddots & 0 \\
0 & \cdots & 0 & h_{L}^{(kj)} & \cdots & h_0^{(kj)} \\
\end{array} \right]
\end{eqnarray*}
\end{small}
$\Fm$ is an FFT matrix with $[\Fm]_{kl}=\exp(-2\pi j \frac{kl}{N} )$
for $k,l=0,\dots,N-1$, and $\xv_k$ denotes the transmit vector of
user $k$ of size $N+L$ subject to the individual power constraint
given by
\begin{equation}\label{PConstraintX}
 \trace(\EE[\xv_k\xv_k^H]) \leq (N+L) P_k
\end{equation}
and $\nv_k\sim\Nc_{\Cc}(\zerov, \Id_N)$ is AWGN. For the primary
user, we consider DFFT-modulated symbols
\begin{equation}\label{x1}
  \xv_1 = \Am \Fm^H \sv_1
\end{equation}
where $\Am$ is a precoding matrix to append the last $L$ entries of $\Fm^H \sv_1$ and $\sv_1$ is a symbol vector of size $N$. For the secondary user, we form the transmit vector by $\xv_2=\Vm \sv_2$ where $\Vm$ is a linear precoder and $\sv_2$ is the symbol vector (whose dimension is be specified later). Our objective is to design the precoder $\Vm$ that generates zero interference, i.e. satisfies the following orthogonal condition
\begin{equation}\label{NullCondition}
    \Tc(\hv^{(21)}) \Vm \sv_2 = \zerov, \;\; \forall \sv_2.
\end{equation}


\section{VFDM}
In this section, we propose a linear Vandermonde precoder that satisfies (\ref{NullCondition}) by exploiting the redundancy $L$ of the cyclic prefix or equivalently the degrees of freedom left by the system. Namely, we let $\Vm$ to be a $(N+L)\times L$ Vandermonde matrix given by

\begin{small}
\begin{eqnarray}
\Vm= \left[ \begin{array}{cccccc}
1 & \cdots & 1 \\
a_1 & \cdots & a_{L} \\
a^2_1 & \cdots & a^2_{L} \\
\vdots & \ddots & \vdots \\
a^{N+L-1}_1 & \cdots & a^{N+L-1}_{L} \\
\end{array} \right] 
\end{eqnarray}
\end{small}

where $\{a_l,\dots,a_L\}$ are the roots of the polynomial $S(z)= \sum_{i=0}^L h_i^{(21)} z^{L-i}$ with $L+1$ coefficients of the channel $\hv^{(21)}$. Since the orthogonality between the precoder and the channel enables two users to transmit simultaneously over the same frequency band, we name this scheme {\it Vandermonde Frequency Division Multiplexing} (VFDM) .
Clearly, the secondary user needs to know perfectly the channel $\hv^{(21)}$ in order to adapt the precoder. This can be done easily assuming that the reciprocity can be exploited under time-division duplexing systems. The resulting transmit vector of the secondary user is given by

\begin{equation}\label{x2}
  \xv_2 =\alpha \Vm\sv_2
\end{equation}

where $\sv_2$ is a symbol vector of size $L$ with covariance $\Sm_2$
and $\alpha$ is determined to satisfy the power constraint (\ref{PConstraintX})
\begin{small}
\[ \alpha = \sqrt{\frac{(N+L)P_2}{\trace(\Vm \Sm_2\Vm^H)}}. \]
\end{small}
The following remarks are in order : 1) Since the channels
$\hv^{(21)}$ and $\hv^{(22)}$ are statistically independent, the probability that 
$\hv^{(21)}$ and $\hv^{(22)}$ have the
same roots is zero. Therefore the secondary user's symbols $\sv_2$ shall be
transmitted reliably; 2) Due to the orthogonality between the
channel and the precoder, the zero interference condition
(\ref{NullCondition}) always holds irrespectively of the secondary
user' input power $P_2$ and its link $\sigma_{2,1}$. This is in
contrast with \cite{jovicic-cognitiveradio-2006} where the zero
interference is satisfied only for the weak interference case, i.e.
$\sigma_{2,1}P_2\leq P_1$ and $\sigma_{1,1}=\sigma_{2,2}=1$; 3) To
the best of our knowledge, the use of a Vandermonde matrix at the
transmitter for interference cancellation has never been proposed. In
\cite{scaglione2000lvm}, the authors proposed a Vandermonde filter
but for a different application.

By substituting (\ref{x1}) and (\ref{x2}) into $\yv_1$, we obtain $N$ parallel channels for the primary user given by

\begin{equation} \label{Y1}
    \yv_1 
    = \Hm_{\rm diag}^{(11)} \sv_1 + \nuv_1
\end{equation}

where $\Hm_{\rm diag}^{(11)}=\diag(H_1^{(11)}, \dots,H_N^{(11)})$ is a diagonal frequency domain channel matrix with i.i.d. entries $H_n^{(11)}\sim\Nc_{\Cc}(0,\sigma_{11})$
and $\nuv_1\sim\Nc_{\CC}(\zerov, \Id)$ is AWGN. The received signal of the secondary user is given by

\begin{equation}\label{Y2}
 \yv_2 
 = \Hm_2 \sv_2 + \Hm_{\rm diag}^{(12)}\sv_1 + \nuv_2
\end{equation}

where we let $\Hm_2=\alpha\Fm \Tc(\hv^{(22)})\Vm$ denote the
overall $N\times L$ channel, $\Hm_{\rm diag}^{(12)}=\diag(H_1^{(12)}, \dots,H_N^{(12)})$ denotes a diagonal frequency domain channel matrix with i.i.d. entries $H_n^{(12)}\sim\Nc_{\Cc}(0,\sigma_{12})$, and $\nuv_2\sim\Nc_{\CC}(\zerov, \Id_N)$ is AWGN.

From (\ref{Y1}) and (\ref{Y2}), we notice that VDFM converts the
frequency-selective interference channel (\ref{OriginalOutput}) into
one-side vector interference channel (or Z interference channel)
where the primary receiver sees interference-free $N$ parallel
channels and the secondary receiver sees the interference from the
primary transmitter. Notice that even for a scalar Gaussian case the
capacity of the one-side Gaussian interference channel is not fully
known \cite{sason2004arr, etkin2007gic}. In this work, we restrict
our receiver to a single user decoding strategy which is clearly suboptimal for the strong interference case $\sigma_{12}>\sigma_{11}$. 


\section{Input Covariance Optimization}\label{sect:InputCovariance}
This section considers the maximization of the achievable rates under the individual power constraints.
First, we consider the primary user. Since the primary user sees $N$ parallel channels (\ref{Y1}), its capacity is maximized by Gaussian input and a diagonal covariance, i.e. $\Sm_1=\diag(p_{1,1},\dots,p_{1,N})$. The rate of the primary user is given by

\begin{equation}
R_1 = \max_{\{p_{1,n}\} }\frac{1}{N} \sum_{n=1}^N \log(1+p_{1n} |H_n^{(11)}|^2)
\end{equation}

with the constraint $\sum_{n=1}^N p_{1,n}\leq N P_1$ \footnote{The power constraint considered here is different from (\ref{PConstraintX}). However, the waterfilling power allocation of (\ref{WaterFilling1}) satisfies (\ref{PConstraintX}) in a long-term under the i.i.d. frequency-domain channels. }. The set of powers can be optimized via a classical waterfilling approach.

\begin{eqnarray}\label{WaterFilling1}
    p_{1,n} = \left[\mu_1 -\frac{1}{|H_n^{(11)}|^2}\right]_+
\end{eqnarray}

where $\mu_1$ is a Lagrangian multiplier that is determined to satisfy $\sum_{n=1}^N p_{1n}\leq N P_1$.

The received signal of the secondary user (\ref{Y2}) when treating the signal from the primary transmitter as noise
is further simplified to
\[\yv_2 = \Hm_2\sv_2 + \etav\]
where $\etav$ denotes the noise plus interference term whose covariance is given by
\[ \Sm_{\eta} =  \Hm_{\rm diag}^{(12)} \Sm_1 {\Hm_{\rm diag}^{(12)}}^H + \Id_N\] 
Under the Gaussian approximation of $\etav$, the rate of the secondary user is maximized by solving

\begin{small}
\begin{eqnarray*}
    \mbox{maximize} &&\frac{1}{N} \log\left|\Id_N + \frac{(N+L)P_2}{\trace(\Vm\Sm_2\Vm^H)}\Gm \Sm_2 \Gm^H \right| \\
    \mbox{subject to} && \trace(\Sm_2) \leq LP_2
\end{eqnarray*}
\end{small}

where we define the effective channel as $\Gm=\Sm_{\eta}^{-1/2}\Hm_2\in \CC^{N\times L}$. Notice that the above problem can be solved with perfect knowledge of the covariance $\Sm_{\eta}$ at the secondary transmitter, which requires the secondary receiver to estimate $\Sm_{\eta}$ during a listening phase and feed it back to its transmitter. The above optimization problem is non-convex since the objective function is neither concave or convex in $\Sm_2$. Nevertheless, we propose a two-step optimization approach that aims at finding the optimal $\Sm_2$ efficiently. The first step consists of diagonalizing the effective channel in order to express the objective function as a function of powers.
We apply singular value decomposition to the effective channel such that $\Gm = \Um_g\Lambdam_g \Pm_g^H$ where
$\Um_g\in\CC^{N\times N}$, $\Pm_g\in\CC^{L\times L}$ are unitary matrices and $\Lambdam_g$ is diagonal with
$r\leq L$ singular values $\{\lambda_{g,l}^{1/2}\}$. Clearly, the optimal $\Sm_2$ should have
the structure $\Pm_g \hat{\Sm}_g\Pm_g^H$ where $\hat{\Sm}=\diag(p_{2,1},\dots,p_{2,r})$ is a
diagonal matrix, irrespectively of the scaling $\trace(\Vm\Sm_2\Vm^H)$. 

For a notation simplicity let us define the signal-to-interference ratio of channel $i$
\begin{small}
\[\SIR_i = (N+L)P_2  c_i\frac{\beta_i p_{2,i}}{\sum_{j=1}^r \beta_j p_{2,j}} \]
\end{small}
where we let $\beta_i\eqdef [\Pm_g^H \Vm^H\Vm \Pm_g]_{i,i}$ and $c_i=\frac{\lambda_{g,i}}{\beta_i}$. By using these notations, it can be shown that the the rate maximization problem reduces to

\begin{small}
\begin{eqnarray}\label{NewProblem} \nonumber
    \mbox{maximize} &&f(\pv_2)= \frac{1}{N}\sum_{i=1}^L \log\left(1+ \SIR_i \right)\\
    \mbox{subject to} && \sum_{i=1}^r p_{2,i} \leq LP_2
\end{eqnarray}
\end{small}

where we let $\pv_2=(p_{2,1},\dots,p_{2,r})$. The second step consists of solving the above power optimization problem. Unfortunately the objective function is not concave in $\{p_{2,i}\}$. Let us first assume that the high SIR approximation is valid for any $i$, i.e.
$\SIR_i\gg 1$ (this is the case for large $(N+L)P_2 c_i$ and in particular when the secondary user's channel is interference-free). Under the high SIR assumption, the function $f$ can be approximated to $J(\pv_2) = \frac{1}{N}\sum_{i=1}^r\log\left(\SIR_i\right)$. 
It is well known that this new function can be transformed into a
concave function through a log change of variable \cite{GP}. Namely let define $\tilde{p_i}=\ln p_i$ (or
$p_i=e^{\tilde{p_i}}$). The new objective function is defined by

\begin{small}
\begin{equation}
    J(\tilde{\pv}_2) = \frac{1}{N}\sum_{i=1}^r (\log(a_i) + \tilde{p}_{2,i})-\frac{r}{N} \log\left(\sum_{j=1}^r\lambda_j e^{\tilde{p}_{2,j}}\right)
\end{equation}
\end{small}

where we let $a_i=(N+L)P_2  c_i\beta_i$. The function $J$ is now concave in $\tilde{\pv}_2$ since the first term is linear and the second term is convex in $\tilde{\pv}_2$. Therefore we solve the KKT conditions which are necessary and sufficient for the optimality. It can be shown that the optimal power allocation reduces to a very simple waterfilling approach given by

\begin{equation}\label{MyWF}
    p_{2,i}^{\star} = \frac{LP_2}{\beta_i\sum_{j=1}^r \frac{1}{\beta_j}}
\end{equation}

which equalizes $\beta_1 p_{2,1}^{\star}=\dots=\beta_r p_{2,r}^{\star}$ and yields $\SIR_i = \frac{(N+L)P_2c_i}{r}$. The resulting objective value would be

\begin{eqnarray*}
f_r = \frac{1}{N} \sum_{i=1}^r \log\left(1+ \frac{(N+L)P_2c_i }{r} \right)
\end{eqnarray*}

It is worth noticing that the high SIR approximation is not necessarily satisfied due to the interference from the primary user and that the optimal strategy should select a subset of channels. One possible heuristic consists of combining the waterfilling based on high SIR approximation with a greedy search. Let us first sort the channels such that

\begin{equation}\label{Sort}
    c_{\pi(1)} \geq c_{\pi(2)} \geq \dots \geq c_{\pi(r)}
\end{equation}

where $\pi$ denotes the permutation. We define the objective value achieved for a subset $\{\pi(1),\dots,\pi(l)\}$ using the waterfilling solution (\ref{MyWF}) with cardinality $l$

\begin{eqnarray}\nonumber
f_l  &=& \frac{1}{N}\sum_{i=1}^l \log\left(1+ \frac{(N+L)P_2c_{\pi(i)}}{l}  \right)
\end{eqnarray}

The greedy procedure consists of computing $f_l$ for $l=1,\dots,r$ and sets the effective number of channels $r^{\star}$ to be the argument maximizing $f_l$.  
As a result, the secondary user achieves the rate given by

\begin{equation}\label{R2}
R_2= \frac{1}{N}\sum_{i=1}^{r^{\star}} \log\left(1+ \frac{(N+L)P_2 c_{\pi(i)}}{r^{\star}}  \right)
\end{equation}

\begin{figure}[ht]
 \centering
 \includegraphics[width=0.4\columnwidth]{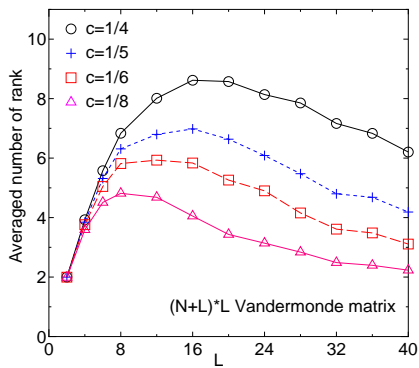}
 \caption{Rank of $(N+L)\times L$ Vandermonde matrix vs. $L$}
 \label{fig:Rank}
\end{figure}

From the rate expression (\ref{R2}), it clearly appears that the
rate of the secondary user (the pre-log factor) depends critically
on the rank $r$ of the overall channel $\Hm_2$, which is determined
by the rank of $\Vm$ since $\Fm, \Tc(\hv^{(22)})$ are full-rank. It
turns out that the rank of $\Vm$ is very sensitive to the amplitude
of the roots $\{a_l\}$ especially for large $N,L$. Although the
roots tend to be on a unit circle as $N,L\rightarrow \infty$ while
keeping $L/N=c$ for some constant
$c>0$ \cite{paper:ibragimovzeitouni}, a few roots outside the unit circle (with $|a_l|>1$) tend to
dominate the rank. In other words for a fixed fraction $c$. Fig. \ref{fig:Rank}
shows the averaged number of ranks of a $5L\times L$ Vandermonde
matrix (corresponding to $c=1/4$) versus $L$. The figure
shows that for a fixed $c$ there is a critical size $L^{\star}$
above which the rank decreases and this size decreases for a larger $N$. This suggests an appropriate choice of
the parameters to provide a satisfactory rate to the
secondary user with VFDM. When the size of the cyclic prefix is larger than
$L^{\star}$, VFDM can be suitably modified so as to boost the
secondary user's rate at the price of increased interference (or
reduced rate) at the primary user. This can be done for example by
normalizing the roots computed by the channel or by selecting $L$
columns from $(N+L)\times (N+L)$ FFT matrix. The design of the
Vandermonde precoder by taking into account the tradeoff between the
interference reduction and the achievable rate is beyond the scope
of this paper and will be studied in a separate paper
\cite{paper:sampaiokobayashi2} using the theory of Random
Vandermonde Matrices
\cite{ryandebbah:vandermonde1,ryandebbah:vandermonde2}.

\section{Numerical examples}
This section provides some numerical examples to illustrate the performance of VFDM with the proposed power allocation. 
Inspired by 802.11a \cite{standard802.11}, we let $N=64, L=16$. 

\begin{figure}[ht]
 \centering
 \includegraphics[width=0.4\columnwidth]{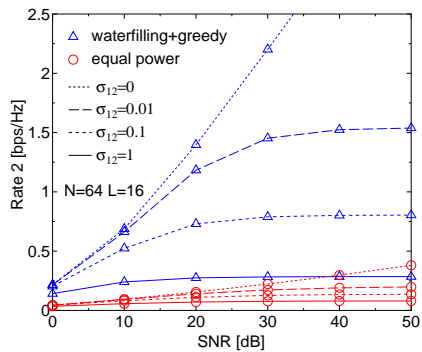}
 \caption{Rate of user 2 vs. SNR}
 \label{fig:Rate2}
\end{figure}

Fig. \ref{fig:Rate2} shows the average rate of the secondary user as a function of SNR $P_1=P_2$ in dB. We let $\sigma_{11}=\sigma_{22}=1$ and vary $\sigma_{12}=1.0,0.1, 0.01, 0.0$ for the link $\hv^{(12)}$. Notice  $\sigma_{12}=0$ corresponds to a special case of no interference. We compare the VFDM performance with equal power allocation $\Sm_2=P\Id_L$ and with the waterfilling power allocation enhanced by a greedy search.
We observe a significant gain by our waterfilling approach and this gain becomes even significant as the interference decreases. This example clearly shows that the appropriate design of the secondary transmitter's input covariance is essential for VFDM. Although not plotted here, the optimization of the primary user's input covariance has a negligible impact on the rates of two users. Finally, it can be shown that the secondary user's rate becomes bounded as $P\rightarrow\infty$ for any $\sigma_{12}>0$ independently of the input covariance.

\begin{figure}[ht]
 \centering
 \includegraphics[width=0.4\columnwidth]{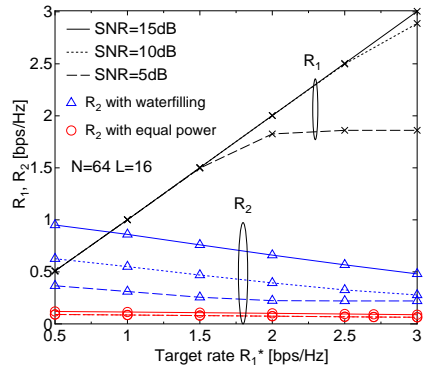}
 \caption{$R_1,R_2$ vs. a target rate $R_1^{\star}$}
 \label{fig:TargetRate}
\end{figure}

Next we consider the scenario where the system imposes a target rate $R_1^{\star}$ to the primary user and the primary transmitter minimizes its power such that $R_1^{\star}$ is achieved. The system sets the transmit power to its maximum $P_1$ if the rate is infeasible. Fig. \ref{fig:TargetRate} shows the achievable rates of both users as a function of the target rate $R_1^{\star}$ in bps/Hz with $P_1=P_2=10$ dB. Again we observe a significant gain due to the appropriate design of the secondary input covariance.

We wish to conclude this section with a simple numerical example inspired by the IEEE 802.11a setting \cite{standard802.11}, showing that VFDM with the appropriate input covariance design enables the secondary user to achieve a considerable rate while guaranteeing the primary user to achieve its target rate over interference-free channels. For example, for the target rate of $R_1=2.7, 1.8$ [bps/Hz] that yields the two highest rates of $54, 36$ [Mbps] over a frequency band of 20MHz, the secondary user can achieves 6.06, 8.44 [Mbps] respectively with operating SNR of 10 dB.

\section{Aknowledgements}
This work was partially supported by Alcatel-Lucent.

\bibliographystyle{unsrt}
\begin{footnotesize}
\bibliography{vfdm}
\end{footnotesize}

\end{document}

%% file: macros.tex
\newfont{\bbb}{msbm10 scaled 500}

\newfont{\bb}{msbm10 scaled 1100}
\newcommand{\CC}{\mbox{\bb C}}

\newcommand{\EE}{\mbox{\bb E}}

\newcommand{\hv}{{\bf h}}

\newcommand{\nv}{{\bf n}}

\newcommand{\pv}{{\bf p}}

\newcommand{\sv}{{\bf s}}

\newcommand{\xv}{{\bf x}}
\newcommand{\yv}{{\bf y}}

\newcommand{\zerov}{{\bf 0}}

\newcommand{\Am}{{\bf A}}

\newcommand{\Fm}{{\bf F}}
\newcommand{\Gm}{{\bf G}}
\newcommand{\Hm}{{\bf H}}
\newcommand{\Id}{{\bf I}}

\newcommand{\Pm}{{\bf P}}

\newcommand{\Sm}{{\bf S}}

\newcommand{\Um}{{\bf U}}

\newcommand{\Vm}{{\bf V}}

\newcommand{\Cc}{{\cal C}}

\newcommand{\Nc}{{\cal N}}

\newcommand{\Tc}{{\cal T}}

\newcommand{\etav}{\hbox{\boldmath$\eta$}}

\newcommand{\nuv}{\hbox{\boldmath$\nu$}}

\newcommand{\Lambdam}{\hbox{\boldmath$\Lambda$}}

\newcommand{\diag}{{\hbox{diag}}}

\newcommand{\trace}{{\hbox{tr}}}

\newcommand{\SIR}{{\sf SIR}}

\newcommand{\eqdef}{\stackrel{\Delta}{=}}